\documentclass[twocolumn,preprintnumbers,amsmath,amssymb,superscriptaddress,subeqn,prl]{revtex4-1}

\usepackage{amsmath}
\usepackage{stmaryrd}
\usepackage[pdftex]{graphicx} % for pdf figures
\usepackage{url}      % use url's in references
\usepackage{times}    % font
\usepackage{float}

\begin{document}
\title{Quasiparticle spectra from a non-empirical optimally-tuned \\range-separated hybrid density functional}

\author{Sivan Refaely-Abramson}
\affiliation{Department of Materials and Interfaces, Weizmann Institute of Science, Rehovoth 76100, Israel}

\author{Sahar Sharifzadeh}
\affiliation{Molecular Foundry, Lawrence Berkeley National Laboratory, Berkeley, CA 94720}

\author{Niranjan Govind}
\affiliation{William R. Wiley Environmental Molecular Sciences Laboratory, Pacific Northwest National Laboratory,
Richland, WA 99352, USA}

\author{Jochen Autschbach}
\affiliation{Department of Chemistry, University at Buffalo, State University of New York, Buffalo, NY 14260-3000, USA}

\author{ Jeffrey B. Neaton}
\affiliation{Molecular Foundry, Lawrence Berkeley National Laboratory, Berkeley, CA 94720}

\author{Roi Baer}
\affiliation{Fritz Haber Center for Molecular Dynamics, Institute of Chemistry, Hebrew University, Jerusalem 91904, Israel}

\author{Leeor Kronik}
\affiliation{Department of Materials and Interfaces, Weizmann Institute of Science, Rehovoth 76100, Israel}

\date{\today}

\begin{abstract}
We present a method for obtaining outer valence quasiparticle excitation energies from a DFT-based calculation, with accuracy that is comparable to that of many-body perturbation theory within the GW approximation. The approach uses a range-separated hybrid density functional, with asymptotically exact and short-range fractional Fock exchange. The functional contains two parameters - the range separation and the short-range Fock fraction. Both are determined {\it non-empirically}, per system, based on satisfaction of exact physical constraints for the ionization potential and many-electron self-interaction, respectively. The accuracy of the method is demonstrated on four important benchmark organic molecules: perylene, pentacene, 3,4,9,10-perylene-tetracarboxylic-dianydride (PTCDA) and 1,4,5,8-naphthalene-tetracarboxylic dianhydride (NTCDA). We envision that for finite systems the approach could provide an inexpensive alternative to GW, opening the door to the study of presently out of reach large-scale systems. \end{abstract}

\maketitle

Development of a non-empirical theory for quantitative electronic structure calculations, which combines predictive power with computational simplicity, is a long-standing challenge for molecular and solid-state physics \cite{Harrison, Martin}. Presently, many-body perturbation theory within the GW approximation \cite{Hedin1965, Aryasetiawan1998, Onida2002} is widely considered to be the first principles approach that provides the best balance between accuracy and computational tractability. This approach is couched in a formally rigorous theory for quasiparticle excitations and has been shown to provide remarkably quantitative predictions for the electronic structure of a wide variety of molecular, solid-state, and low-dimensional systems (see, e.g., \cite{Hybertsen1986, Aryasetiawan1998, Aulbur, Onida2002, Spataru2004, *Yang2007, Blase2011, Giantomassi2011}).

Unfortunately, present-day GW calculations are still significantly limited in system size and complexity. They can also be challenging to converge \cite{Shih2010, Frierdrich2011, Sharifzadeh2012}. Therefore, it is common practice to rely instead on density functional theory (DFT) \cite{DreizlerGross}, which is much simpler computationally. However, this comes at a significant cost in accuracy. Solutions of the Kohn-Sham equation (in either its original \cite{Kohn1965} or generalized \cite{Seidl1996} form) generally do not rigorously correspond to quasiparticle energies and orbitals. Practical DFT calculations can still be, and often are, successful because occupied DFT eigenvalues can, in principle, serve as good \textit{approximations} to removal energies of energetically high-lying occupied orbitals \cite{Kummel2008, Hybertsen1986, Chong2002, Gritsenko2002, Baer2009}. Even so, two major problems remain \cite{Kummel2008}. First, it is often found that the eigenvalue spectrum can depend strongly, and even qualitatively, on the choice of the approximate density functional. Second, the energies of the highest occupied molecular orbital (HOMO) and the lowest unoccupied molecular orbital (LUMO) typically do not correspond to the ionization potential and electron affinity, respectively.

In this Letter, we show that DFT-based calculations, in which outer-valence orbitals do represent quasiparticle excitations, are in fact possible, opening the door to inexpensive prediction of quasiparticle excitations. Our approach is based on a range-separated hybrid density functional, which is optimally tuned to obey Koopmans' (ionization potential) theorem and to minimize many-electron self-interaction errors, without any recourse to empiricism.

Recently, Stein \textit{et al.} \cite{Stein2010} suggested a new method for predicting the fundamental gap of finite systems from generalized Kohn-Sham HOMO and LUMO eigenvalues, based on a non-empirical optimally-tuned range-separated hybrid (OT-RSH) functional. In an RSH functional, the Coulomb repulsion is partitioned into a short-range (SR) and a long-range (LR) part, such that the LR exchange is treated with a Fock operator whereas the SR exchange is treated using (semi-)local exchange \cite{Leininger1997}. The range-separation parameter, $\gamma$, is optimally-tuned by demanding that the DFT version of Koopmans' theorem (ionization potential theorem) be obeyed, i.e., by determining $\gamma$, per-system, such that the HOMO eigenvalues of the neutral and anionic system are as close as possible to the ionization potential and electron affinity of the neutral, respectively, making the gap contribution of a derivative discontinuity \cite{Sham1983, *Perdew1983, Kummel2008,Kronik2012} negligible. Refaely-Abramson \textit{et al.} have shown the efficacy of this approach, using the optimally-tuned RSH, denoted here as OT-$\gamma$RSH, for a range of organic molecules of relevance to photovoltaics \cite{Refaely-Abramson2011}. This results in HOMO and LUMO levels on par with the GW ones. It can perhaps be hoped, then, that the outer-valence occupied orbitals of finite systems could also be well described with this tuned functional.

\begin{figure*}
%\begin{center}
\includegraphics[scale=0.7]{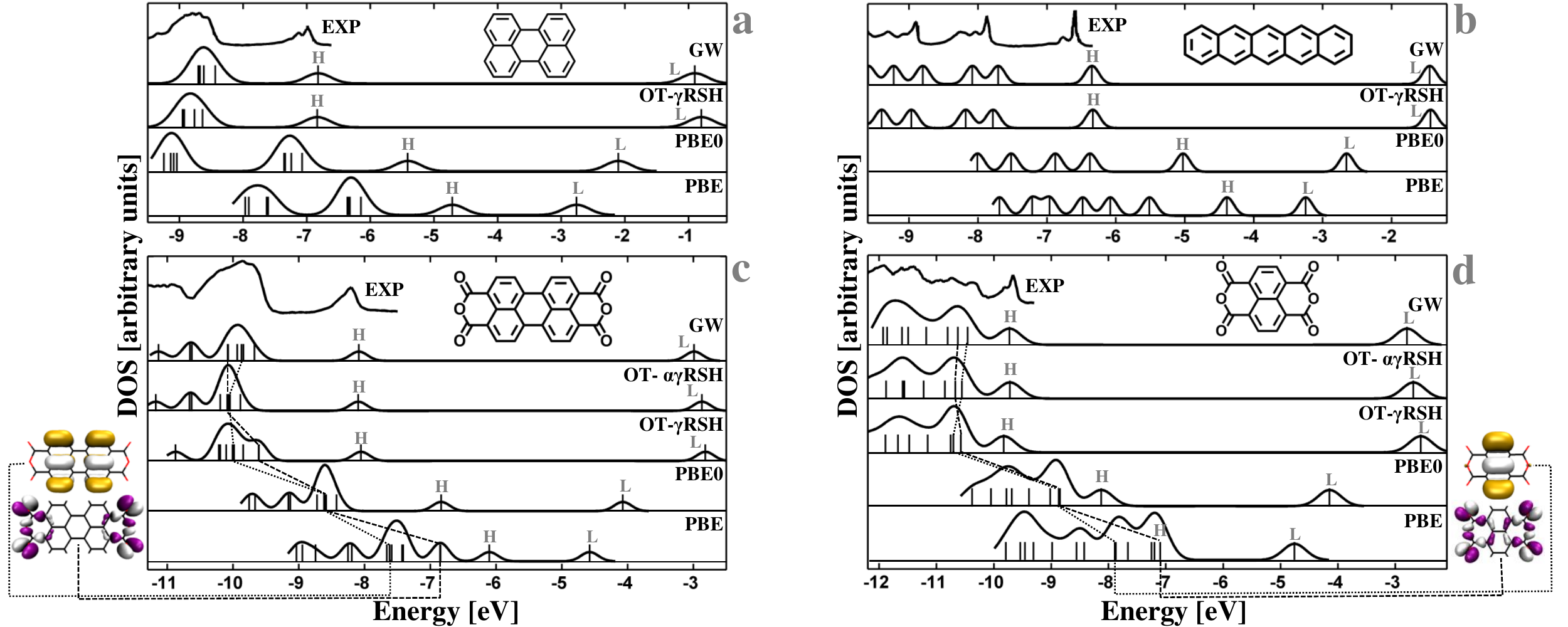}
\centering
\caption{(Color online) Outer valence eigenvalue spectra of (a) Perylene, (b) Pentacene, (c) PTCDA, (d) NTCDA, as obtained from DFT calculations using the PBE, PBE0, and OT-$\gamma$RSH, with additional OT-$\alpha \gamma$RSH results for PTCDA and NTCDA, compared with GW eigenvalue spectra and with experimental gas phase photoemission spectra \cite{PES_fig1}. H and L denote HOMO and LUMO, respectively. For PTCDA and NTCDA, representative localized and delocalized orbitals are presented. All computational spectra have been broadened by convolution with a gaussian to facilitate comparison with experiment.}
\label{fig1}
%\end{center}
\end{figure*}

To test this notion, Fig.~\ref{fig1}(a,b) presents a comparison of experimental photoemission spectra and theoretical DFT and GW eigenvalue spectra \cite{G1W1} for perylene and pentacene. Here and throughout, DFT eigenvalue spectra were performed using QChem \cite{QCHEM} with the cc-PVTZ basis set \cite{Dunning1989} and GW calculations were performed using the BerkeleyGW code \cite{Deslippe2012} (details are given in the supplementary information, SI).
For both molecules, the GW spectrum agrees well with experiment. The DFT spectrum constructed using the generalized gradient approximation (GGA) in its Perdew-Burke-Ernzerhof (PBE) form \cite{PBE1996} does not agree with GW, but is primarily rigidly shifted from it, owing to the missing derivative-discontinuity in the exchange-correlation functional \cite{Perdew1983}. Hybrid functionals, e.g., PBE0\cite{Adamo1999} (which is based on the PBE functional  but with 25\% of the GGA exchange replaced by Fock exchange) mitigate (but do not solve) the derivative discontinuity problem \cite{Kummel2008,Kronik2012} and consequently improve the spectrum by shifting (and slightly stretching) it. Still, the PBE0 spectrum exhibits a significant rigid shift. Conversely, the OT-$\gamma$RSH is in quantitative agreement with GW and experiment (average unsigned error of $\sim$0.2 eV over a range of $\sim$3 eV below the HOMO), as hoped for.

%when compared to GW spectra and to experiment, with an The PBE0 hybrid functional \cite{Adamo1999} spectrum, also plotted for comparison, predicts well-ordered but roughly shifted orbitals, as discussed below. For deeper low-lying states, the situation is more complicated and OT-$\gamma$RSH accuracy is strongly dependent on the orbital shape.

Unfortunately, this simple idea is not sufficient for more complex molecules, such as 3,4,9,10-perylene-tetracarboxylic-dianydride (PTCDA) and 1,4,5,8-naphthalene-tetracarboxylic dianhydride (NTCDA) (Fig.~\ref{fig1}(c,d)), where
GW spectra agree with experiment but the OT-$\gamma$RSH spectra present serious deviations from GW in both orbital position and ordering. These molecules have been chosen because, while still reasonably simple, they exhibit a mixture of localized (on the anhydride side groups) and delocalized (on the perylene or naphthalene core) outer valence orbitals, as shown in Fig.~\ref{fig1}(c,d). Dori et al.\ \cite{Dori2006} pointed out that this causes the PBE spectrum of PTCDA to be in poor agreement with GW even after a rigid shift, as also shown in Fig.~\ref{fig1}(c). The spectral distortions result primarily from spurious positive energy shifts of the localized orbitals, leading to the conjecture that they reflect significant self-interaction errors (SIE)\cite{Dori2006}. K\"{o}rzd\"{o}rfer \textit{et al.} \cite{Korzdorfer2009} showed that NTCDA has a similar problem, as can be seen in Fig.~\ref{fig1}(d), and proved the conjecture by quantifying the per-orbital SIE for both molecules. Furthermore, they showed that self-interaction-corrected calculations, within a generalized optimized effective potential scheme, provide a non-empirical route for obtaining agreement with experiment, up to a rigid shift of both HOMO and LUMO (and possibly some mild stretching).

For both PTCDA \cite{Dori2006} and NTCDA \cite{Korzdorfer2009}, a different non-empirical route for improvement of the eigenvalue spectral shape is the use of the above-mentioned PBE0 hybrid functional, which possesses a fixed fraction of exact exchange, as shown in Fig.~\ref{fig1}(c,d). Indeed, K\"{o}rzd\"{o}rfer and K\"{u}mmel \cite{Korzdorfer2010} showed that such hybrids can incorporate an important part of the first order correction of the Kohn-Sham eigenvalues when used in a generalized Kohn-Sham way, i.e., with non-local Fock exchange. However, recent orbital tomography experiments \cite{Dauth2011,*Puschnig2011} showed that the orbital ordering in hybrids can still be wrong. This is also reflected in Fig.~\ref{fig1}c, and is likely a consequence of the hybrid functional not being self-interaction free. In addition, conventional hybrid functionals do not resolve the ionization potential and fundamental gap problem. Thus, our goal is to provide a generalized Kohn-Sham scheme that does yield the correct excitation thresholds and is also self-interaction free. We now show that this is an achievable goal.

The results of Fig.~\ref{fig1} suggest that further improvement could be obtained by {\it combining} a fraction of SR Fock exchange that would improve the description of the localized orbitals, with LR Fock exchange that is essential for gap prediction. Such generalization of the RSH scheme has in fact been suggested by Yanai \textit{et al.} \cite{Yanai2004}, who partitioned the Coulomb operator according to
\begin{equation}	
\frac{1}{r}=\frac{\alpha+\beta\mathrm{erf}(\gamma r)}{r}+\frac{1-[\alpha+\beta\mathrm{erf}(\gamma r)]}{r}. \label{GOT-RSH}
\end{equation}
Note that this form reduces to that of a conventional hybrid functional with the choice $\beta=0$ (for PBE0, $\alpha=0.25$) and to a "pure LR" RSH for $\alpha=0, \beta=1$.

The parameters $\alpha$, $\beta$, and $\gamma$ can be determined as universal parameters semi-empirically, as done for all three of them in the CAM-B3LYP functional of Yanai \textit{et al.} \cite{Yanai2004} However, as discussed in, e.g., refs. \cite{Stein2010, Refaely-Abramson2011, KorzdorferSears2011, SrebroOptical2012, Kronik2012, Phillips2012}, no set of fixed values is universally useful for spectroscopy. Instead, we pursue the optimal tuning strategy, where we determine all three parameters from first principles, per-system, based on satisfaction of physical constraints.

First, following ref.\ \cite{Rohrdanz2008}, we insist on $\alpha + \beta =1$. This choice guarantees that full Fock exchange is obtained asymptotically, which enforces the correct asymptotic potential. This, in turn, is essential for retaining accurate gap predictions \cite{Kronik2012}. As in Refs. \cite{Vydrov2006, Rohrdanz2009}, we shall use PBE-based semi-local exchange and correlation components. However, in these articles $\alpha$ was taken as a constant of 0 and 0.2, respectively, with $\gamma$ a universal empirically determined constant. We shall seek to optimize both $\alpha$ and $\gamma$ non-empirically, based on additional constraints.

Per each choice of $\alpha$, $\gamma$ can be determined from first principles by enforcing the ionization potential (Koopmans') theorem, i.e., by choosing $\gamma$ such that the HOMO eigenvalue is as close as possible to the ionization potential \cite{Perdew1982, Almbladh1985, *Perdew1997, *Levy1984}. In principle, this exact condition should be obeyed for any stable ionic state of the molecule. Therefore, one can seek $\gamma$ that best satisfies (say, in the least squares sense) multiple ionization potential conditions, by minimizing a target function $J(\gamma)$ of the form:
\begin{equation}
J^2(\gamma;\alpha)=\sum_{i}(\varepsilon^{\gamma;\alpha}_{H(N+i)}+IP^{\gamma;\alpha}(N+i))^2, \label{JPHL}
\end{equation}
where $\varepsilon_{H(N+i)}$ is the HOMO eigenvalue of the $N+i$ electrons system, $N$ being the number of electrons in the neutral system and $i$ an integer representing electrons added or removed from it, with $IP(N+i)$ the corresponding ionization potential, calculated from energy differences.
In previous work that emphasized accurate gap prediction \cite{Stein2010, Refaely-Abramson2011,KorzdorferSears2011,Kronik2012,Phillips2012}, $\gamma$ was chosen so as to satisfy this condition as closely as possible for both neutral and anion (i.e., $i=$0 and 1 in Eq. \ref{JPHL}), so as to obtain both the ionization potential and the electron affinity (the latter being equal to the ionization potential of the anion). Here, this does not suffice, as $\gamma$ must also reflect a balance of SR and LR exchange appropriate for the treatment of localized states. Because with OT-$\gamma$RSH the highest localized orbital is HOMO-1, for both PTCDA and NTCDA, we additionally impose an ionization potential condition for the cation, i.e., $i$=-1, 0, and 1 in Eq. \ref{JPHL} (see SI for additional details).

\begin{figure}
\includegraphics[scale=0.35]{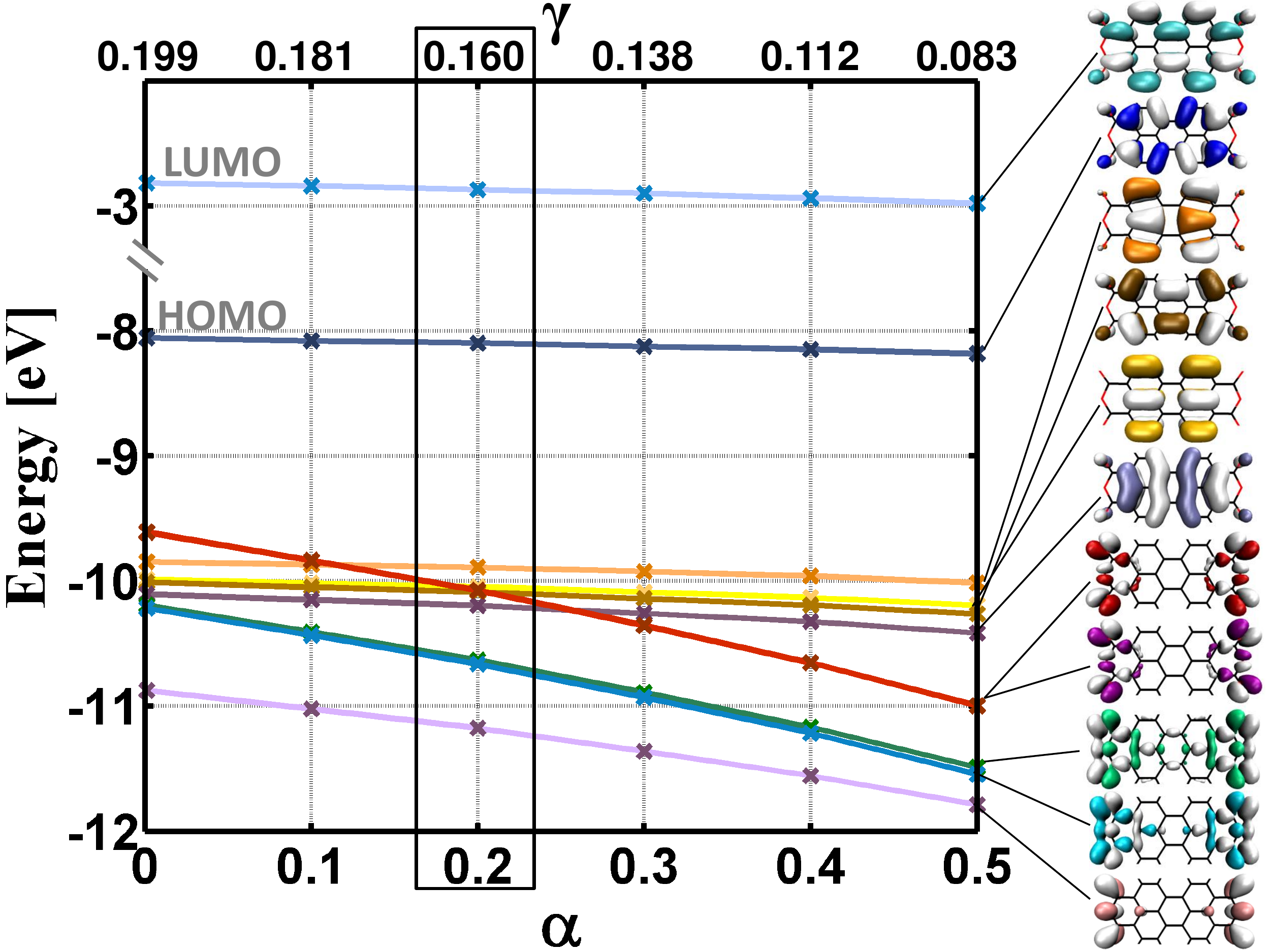}
\caption{(Color online) Eigenvalue energy as a function of the short-range Fock fraction, $\alpha$, with the optimal value of the range-separation parameter, $\gamma$ (Bohr$^{-1}$) deduced for each choice of $\alpha$, for PTCDA. HOMO and LUMO levels are marked. Rectangle denotes the optimal $\alpha$, determined from the minimization procedure described in the text.}
\label{fig2}
\end{figure}

The remaining question, then, is how to determine the SR Fock exchange fraction, $\alpha$. To understand the effect of $\alpha$ on the spectrum, Fig.~\ref{fig2} shows the outer valence eigenvalues as a function of $\alpha$, for the example of PTCDA (similar results for all other molecules are given in the SI). For each choice of $\alpha$, the optimal value of $\gamma$, determined by employing Eq. \ref{JPHL} with a triple summation, has been used and is also shown.
Several important trends can be distinguished immediately. First, as $\alpha$ increases, the optimized $\gamma$ decreases. This is reasonable: the range above which the exchange is dominated by its LR contribution roughly corresponds to $1/\gamma$, and the extent of LR Fock corrections should decrease with increasing SR Fock contributions. Second, for $\alpha$ between 0 and 0.5, $\gamma$ tuning is successful throughout in maintaining a HOMO-LUMO gap that is constant to within $\sim$0.05 eV and is in excellent agreement with GW.  Larger $\alpha$ values are not given because for too large $\alpha$ determining a corresponding $\gamma$ that obeys Koopmans' theorem to a meaningful accuracy is no longer possible. This makes sense, because there is a limit to the extent of SR Fock exchange that can be used while still maintaining compatibility with a semi-local correlation expression \cite{Kummel2008}.

Third, Fig.~\ref{fig2} clearly exposes the different behavior of the two types of orbitals present in the outer valence region. Eigenvalues corresponding to delocalized orbitals (on the perylene core) are essentially indifferent to the choice of $\alpha$ (to within a mean value of 0.2 eV). Conversely, all anhydride-localized orbitals are highly sensitive to $\alpha$. As an example, the doubly-degenerate orbital, that is HOMO-1/2 for $\alpha$=0, is HOMO-5/6 for $\alpha$=0.5, and it changes in energy from $\sim$-1.5 eV to $\sim$-2.8 eV. A similar picture emerges for NTCDA. For perylene and pentacene, all outer valence orbitals within $\sim$3 eV below the HOMO are delocalized and the spectrum is largely independent of $\alpha$ (see SI for details), which explains the success of the SR-exchange-free OT-$\gamma$RSH functional for these molecules (Fig.\ 1(a,b)). Deeper lying orbitals possess different degrees of localization and are outside the scope of this work.

How to choose an optimal $\alpha$, then, without empiricism? Recently, Srebro and Autschbach \cite{Srebro2012} suggested that it can be obtained by insisting on an additional property satisfied by exact DFT, namely, that for ensemble states described by a fractional number of electrons the total energy versus particle number curve must be piecewise linear \cite{Perdew1982}. They used such tuning to obtain an accurate CuCl electric field gradient. We stipulate that satisfaction of the piecewise linearity constraint is important for spectroscopy as well, for two reasons. First, Yang and co-workers have emphasized the importance of linear segments for accurate gap prediction \cite{Mori-Sanchez2008, *Zheng2011, Cohen2012}. Second, enforcing piecewise linearity has been shown to be essential for the accurate spectroscopy of localized states \cite{Lany2009, Dabo2010} and deviation from this condition is in fact often dubbed a "many-electron SIE" or a "delocalization error" \cite{Vydrov2007, Cohen2012}.

\begin{figure}[h]
\includegraphics[scale=0.28]{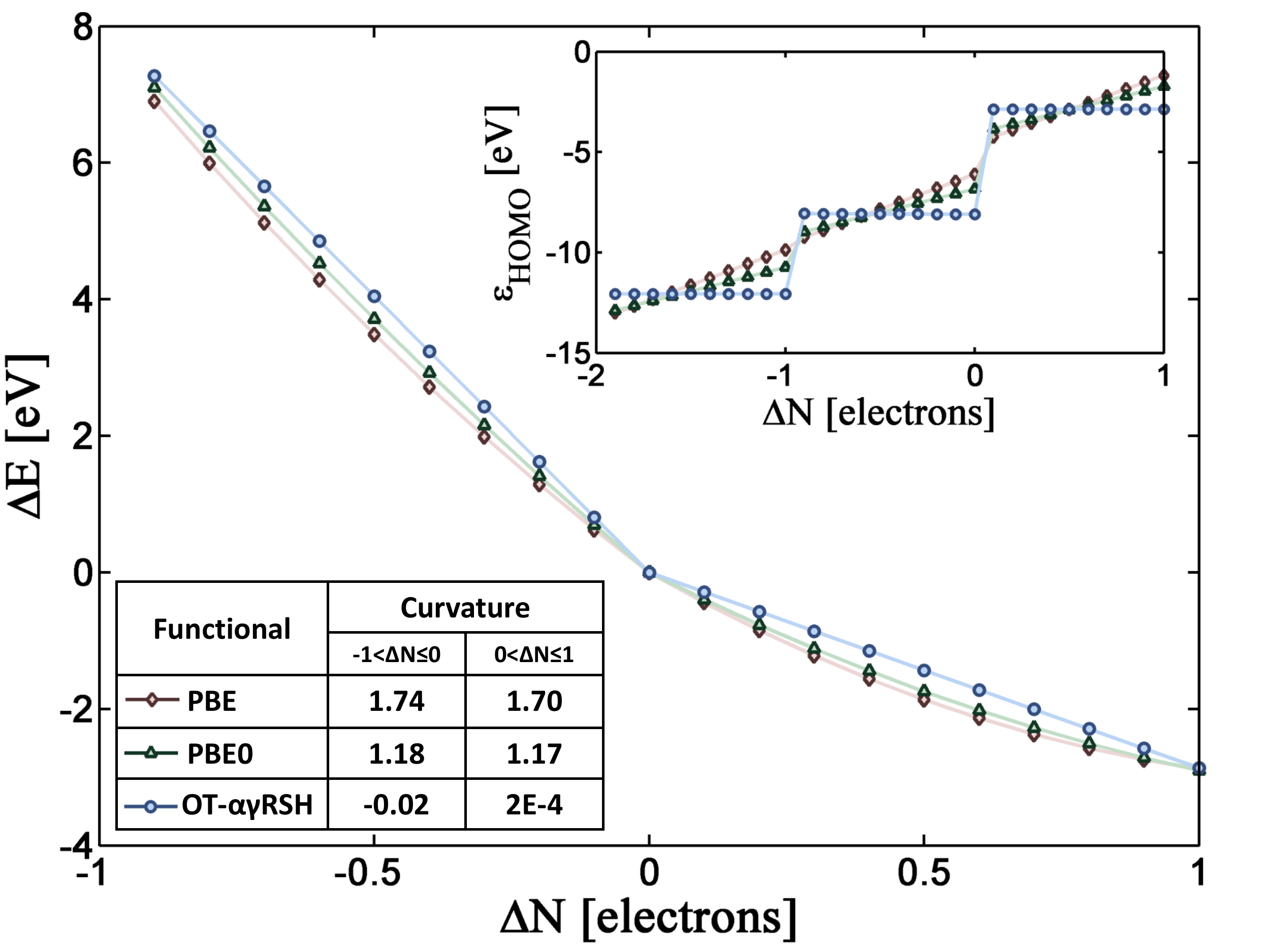}
\caption{(Color online) Deviation of total energy from that of the neutral molecule, $\Delta E$, and HOMO eigenvalue, $\varepsilon_{HOMO}$ (inset), as a function of the fractional deviation of the number of electrons from that of the neutral molecule, $\Delta N$, computed for PTCDA using PBE, PBE0, and OT-$\alpha \gamma$RSH ($\alpha$=0.2, $\gamma$=0.160 Bohr$^{-1}$). The table shows the curvature of each functional, in eV, obtained from fitting the $\Delta E$($\Delta N$) curve to a second order polynomial.}
\label{fig3}
\end{figure}

Several groups have already shown that for a well-constructed RSH functional, curves of the energy as a functional of the fractional number of electrons, for the [N-1,N] and [N, N+1] segments, are much more linear than those obtained with conventional functionals, even in the absence of optimal tuning \cite{Cohen2008, Vydrov2007, Tsuneda2010, SrebroOptical2012, Srebro2012, Kronik2012}. We have performed fractional electron calculations for our benchmark molecules using NWChem \cite{NWCHEM} with the same basis set as above. As shown in Fig.~\ref{fig3} using PTCDA as an example, the above findings apply here as well: Whereas PBE and PBE0 exhibit a notable deviation from linearity, for an OT-RSH functional the deviation from linearity is too small to be detected by the naked eye. An alternative approach to assessing segment linearity, which is more directly relevant to spectroscopy, is to consider the dependence of the eigenvalues on the fractional number of electrons \cite{Vydrov2007}, as shown in the inset of Fig.~\ref{fig3} for [N-2, N+1]. If the linear-segment constraint is satisfied, the HOMO eigenvalue should be constant between integer electron values, owing to Janak's theorem \cite{Janak1978}. Again, only the OT-RSH functional obeys this requirement closely enough. The optimal value of $\alpha$ can thus be obtained by choosing the $\alpha$ (and therefore the corresponding $\gamma$) that minimizes the three curvatures of the $\Delta E$($\Delta N$) curve for $-2<\Delta N \le 1$. It is found to be 0.2 for both PTCDA and NTCDA.

%Deviation from linearity can be quantified, therefore, either by the second derivative of the curves in Fig.~\ref{fig3} or the first derivative of the curves in the inset of Fig.~\ref{fig3}.

Satisfactorily, we find that for both molecules the orbital ordering for the optimal $\alpha$ is similar to the GW ordering (with the exception of PTCDA HOMO-5 orbital which is slightly misplaced), without any need for level shifting: OT-$\alpha \gamma$RSH eigenvalues of PTCDA delocalized states deviate from the same in GW by $\sim$0.18 eV (the largest deviation being 0.25 eV). For localized states, this deviation is $\sim$0.01 eV. For NTCDA, the average deviation of all states from GW is $\sim$0.07 eV. These numbers are well within the accepted accuracy of either calculation. Indeed, the OT-$\alpha \gamma$RSH spectrum, also shown in Fig.~\ref{fig1}, agrees extremely well with the GW one for both absolute HOMO and LUMO positions and the quasiparticle spectrum of filled states, for all examined molecules.

In conclusion, we demonstrated, using four important benchmark molecules, that DFT-based calculations can reach an accuracy that is comparable to that of GW calculations. This was achieved by using a PBE-based range-separated hybrid density functional, with asymptotically exact and short-range fractional Fock exchange. Importantly, both range-separation and Fock fraction are determined non-empirically, based on satisfaction of exact constraints for the ionization potential and many-electron self-interaction error, respectively, resulting in full predictive power for the outer valence electronic structure. We envision that the approach could be useful directly as a low-cost alternative to GW that offers good accuracy for both frontier and non-frontier quasiparticle excitation energies. Additionally, because perturbative "one-shot" G$_0$W$_0$ is known to be sensitive to the DFT starting point \cite{Rinke2008, Blase2011, Marom2011}, our approach provides a novel optimal starting point for subsequent GW calculations.

We thank Stpehan K\"ummel (Bayreuth) for illuminating discussions. Work was supported by the European Research council, the Israel Science Foundation, the United States-Israel Binational Science Foundation, the National Science Foundation, the Molecular Foundry, the Network for Computational Nanotechnology, the U.S. Department of Energy, and the EMSL- a national scientific user facility sponsored by the U.S. Department of Energy's Office of Biological and Environmental Research and located at Pacific Northwest National Laboratory (PNNL). We thank the National Energy Research Scientific Computing center for computational resources.

\bibliographystyle{apsrev4-1}
\bibliography{PTCDA_BIB}

\end{document}